\documentclass[9pt,twocolumn,twoside]{osajnl}
\journal{ol} % Choose journal (ao, aop, josaa, josab, ol, optica, pr)

\setboolean{shortarticle}{true}
\usepackage[markup=underlined]{changes}
\makeatletter
\@namedef{Changes@AuthorColor}{red}
\colorlet{Changes@Color}{red}
\makeatother
\usepackage{bm}

\title{Inverse design of broadband and lossless
topological photonic crystal waveguide modes}

\author[1]{Eric Nussbaum}
\author[1]{Erik Sauer }
\author[1]{Stephen Hughes}
\affil[1]{Department of Physics, Engineering Physics and Astronomy,
Queen's University, Kingston, ON K7L 3N6, Canada}

\begin{abstract}
Topological photonic crystal waveguides can create edge states that may be more robust against fabrication disorder, and can
yield propagation modes below the light line. We present a fully three-dimensional method to modify state-of-the-art designs
to achieve a significant bandwidth improvement for lossless propagation. Starting from an initial design with a normalized bandwidth of 7.5$\%$ (13.4 THz), the modification gives more than 100\% bandwidth improvement to  16.2$\%$ (28.0 THz). This new design is obtained using automatic differentiation enabled inverse design and a guided mode expansion technique to efficiently calculate the band structure and edge state modes.
\end{abstract}

\begin{document}

\maketitle
\thispagestyle{empty}

\section{Introduction}

Photonic crystal slab (PCS) waveguides can slow down light and even stop light (theoretically) in sub-wavelength dimensions, which can be exploited for enhancing various light-matter interactions for applications in sensing, nonlinear optics and quantum optics~\cite{krauss_why_2008}.
Moreover, when integrated with semiconductor quantum dots, PCS waveguides can enable single photons to be produced
with almost 100\% on-chip radiative emission~\cite{manga_rao_single_2007,PhysRevLett.113.093603,PhysRevX.2.011014}.
However, a significant problem
for exploiting such waveguides
is that they suffer, in many cases dramatically,
from disorder-induced scattering~\cite{hughes_extrinsic_2005},
which is particularly severe
in the slow-light regime. This effect is now well understood theoretically and experimentally~\cite{kuramochi_disorder-induced_2005,patterson_disorder-induced-coherent_2009,patterson_disorder-induced-incoherent_2009,OFaolain:07}, and seems to be a fundamental problem, since the slow-light regime is also the regime for exploiting enhanced light-matter interactions.

In recent years, topological PCS waveguides
have been demonstrated
\cite{barik_two-dimensionally_2016,anderson_unidirectional_2017}, which may suppress disorder-induced
backscattering through ``topological protection.''
Moreover, PCS waveguides also exhibit Bloch modes with local chirality~\cite{young_polarization_2015,sollnerDeterministicPhotonEmitter2015,barik_topological_2018},
which can be used to couple to spin charged quantum dots, manifesting in unidirectional single photon emission.
Unfortunately, many of the topological PCS waveguide modes fall above the light line and are thus intrinsically lossy, with significant propagation losses~\cite{sauerTheoryIntrinsicPropagation2020}.
Very recently, new classes of topological PCS waveguides have been presented,
using the so-called Valley Hall effect
\cite{shalaev_robust_2019, heSilicononinsulatorSlabTopological2019},
which more easily
allow  edge state modes to fall below the light line.
For future applications of these topological PCS structures, one goal is to increase the operation bandwidth for single mode operation below the light line.

In this Letter, we demonstrate 
an efficient way to significantly increase the lossless bandwidth for such devices, by combining inverse design techniques~\cite{piggottInverseDesignDemonstration2015, moleskyInverseDesignNanophotonics2018}
with the semi-analytical approach to computing photonic  band structures known as the guided mode expansion (GME) technique~\cite{andreaniPhotoniccrystalSlabsTriangular2006}. 
Starting with a state-of-the-art design from 
 He {\em et al.}~\cite{heSilicononinsulatorSlabTopological2019}, we show how to increase the bandwidth by more than 110$\%$, yielding 
a significant operation bandwidth of 28.0 THz below the light line, with single mode operation that is compatible with telecom wavelengths. Moreover,
 our methodology can be used to optimize many target figures of merit for PCS structures, in an intuitive and efficient way.

\section{Methodology and Theory}

The photonic band structure of PCS waveguides can be efficiently calculated using the GME method, a semi-analytical method~\cite{andreaniPhotoniccrystalSlabsTriangular2006}
that formulates Maxwell's equation as an eigenvalue problem with a lattice structure matrix representation and the guided modes of the slab, allowing one to accurately calculate both the real and imaginary band structure with fast computational run times. 

In this work, we use a GME {\sc Python} package named {\sc Legume}, from Minkov {\em et al.}~\cite{minkovInverseDesignPhotonic2020, FancomputeLegume2020}.
We then combine this approach, which is already very efficient,
with inverse design techniques, to show how to significantly improve the operational bandwidth of topological waveguide modes that are intrinsically lossless, namely how to significantly increase the bandwidth of single mode operation below the light line.
For virtually all applications with integrated photon devices, and device figures of merit, one desires to increase the operational bandwidth.

We first summarize the salient details of the GME. One can rewrite Maxwell's equations in the frequency domain, to obtain an 
eigenvalue equation in terms of the magnetic field, $\bm H(\bm r)$:
\begin{equation}
    \label{eq:maxwell}
    \bm \nabla \times \left( \frac{1}{\epsilon(\bm r)} \bm \nabla \times \bm H(\bm r) \right)  = \left(\frac{\omega}{c}\right)^2 \bm H(\bm r),
\end{equation}
with the condition $\nabla \cdot \bm H(\bm r) =0$, where $\epsilon(\bm r)$ is the dielectric constant.
To solve~\eqref{eq:maxwell},  the GME method expands the magnetic field into an orthonormal set of basis states as
\begin{equation}
    \label{eq:orthonormal basis states}
    \bm H(\bm r) = \sum_\mu c_\mu \bm H_\mu (\bm r),
\end{equation}
and so~\eqref{eq:maxwell}
can be written as
\begin{equation}
    \label{eq:eigen}
    \sum_\nu \mathcal{H}_{\mu \nu}c_\nu = \frac{\omega^2}{c^2} c_\mu,
\end{equation}
where the elements of the Hermitian matrix $\mathcal{H}_{\mu \nu}$ are defined as:
\begin{equation}
    \label{eq:eigenMatrix}
    \mathcal{H}_{\mu \nu} = \int \cfrac{1}{\epsilon(\bm r)} \left(\nabla \times \bm H^{*}_\mu(\bm r) \right) \cdot \left(\nabla \times \bm H_\nu(\bm r) \right) d \bm r.
\end{equation}

To define an appropriate basis set $\bm H_\mu (\bm r)$, the GME method uses the guided modes of the effective homogeneous slab waveguide, with a dielectric constant taken as the spatial average of the dielectric constant in the slab layer. The guided modes of the effective waveguide depend on a wave vector, which can take any value in the slab plane. The modes of the PCS depend on the wave vector, $\bm k$, which can be restricted to the first Brillouin zone. Thus, only the effective waveguide modes with wave vector $\bm k + \bm G$, where $\bm G$ is a reciprocal lattice vector of the PCS, are included in the basis. The guided mode expansion is then
\begin{equation}
    \label{eq:GME magnetic field}
    \bm H_{\bm k} (\bm r) = \sum_{\bm G, \alpha} c(\bm k + \bm G, \alpha) \bm H_{\bm k + \bm G , \alpha} ^{\rm guided} (\bm r),
\end{equation}
where $\bm H_{\bm k + \bm G , \alpha} ^{\rm guided} (\bm r)$ is a guided mode of the effective waveguide and $\alpha$ is the index of the guided mode~\cite{andreaniPhotoniccrystalSlabsTriangular2006}. Once the magnetic field of a photonic mode is found, the orthonormal electric field eigenmodes can be straightforwardly obtained \cite{andreaniPhotoniccrystalSlabsTriangular2006}.

We next describe the inverse design approach, and how it can be implemented with the GME.
Inverse design treats the design process as an optimization problem. An objective function is written, which accepts a device parameterization and returns an appropriate figure of merit (FOM). This function is then optimized, typically with a gradient-based optimization algorithm. The gradient calculations are normally performed using the adjoint variable method, however this method cannot be easily implemented with the GME technique \cite{minkovInverseDesignPhotonic2020}. Instead, we use automatic differentiation (AD), which allows the gradient of an arbitrarily complex function to be calculated.

The fundamental idea behind AD is that the gradient of a function can composed of the gradients of its subfunctions. When using an AD library, the objective function is defined, and then by tracking sub-function executions, the AD library creates an operator to return the objective function gradient. In this work, we use the AD library {\sc Autograd}, which allows the objective function to be written in normal {\sc Python} code and is compatible with most of the {\sc NumPy} library~\cite{maclaurinHIPSAutograd}. The GME library being used, {\sc Legume}, allows its back end to be set to {\sc Autograd} compatible code, making calculating the objective function gradient straightforward.

\section{Design process and results}

\begin{figure}
    \centering
    \includegraphics[width = 0.8 \columnwidth]{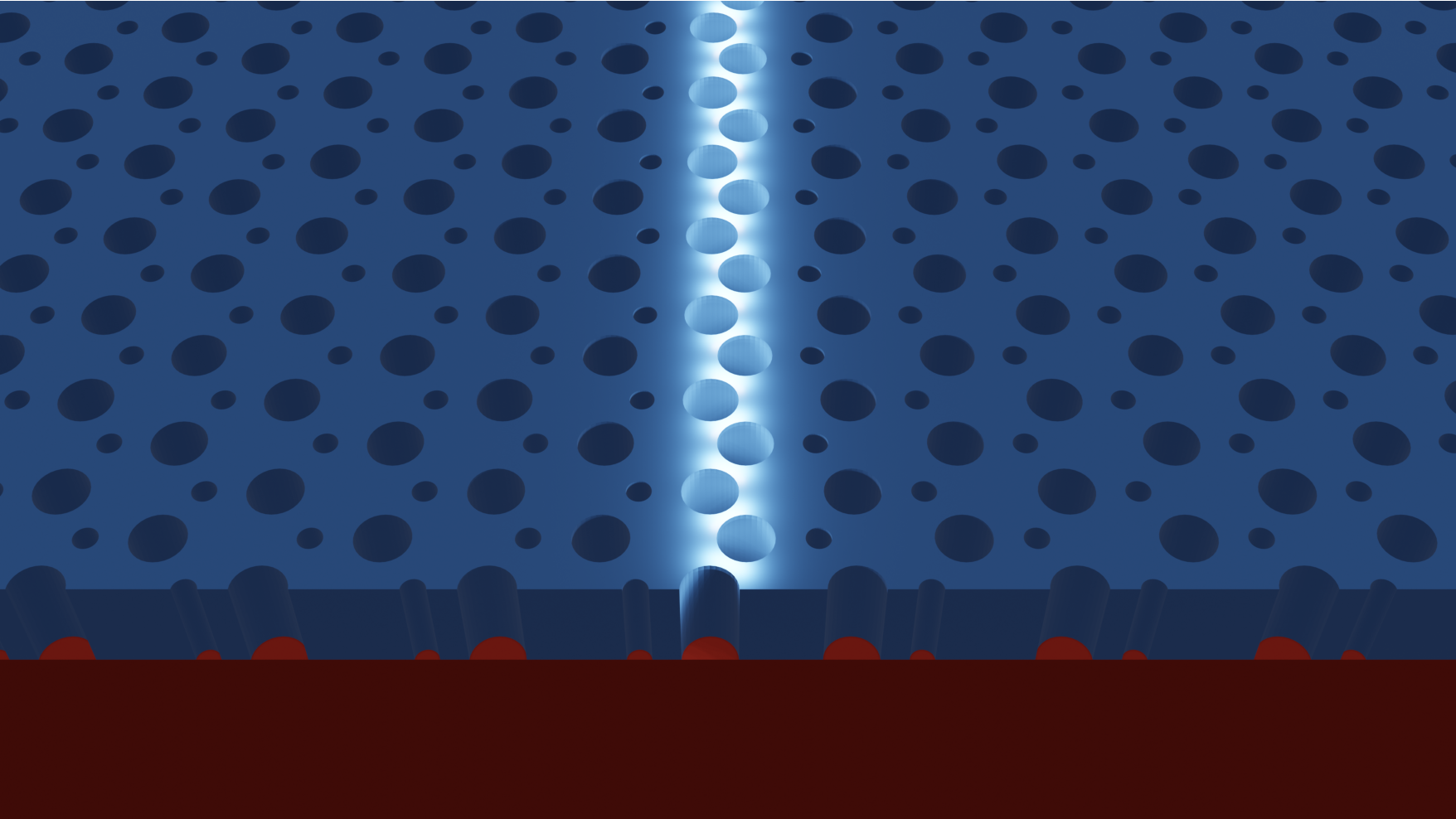}
    \includegraphics[width = 0.8 \columnwidth]{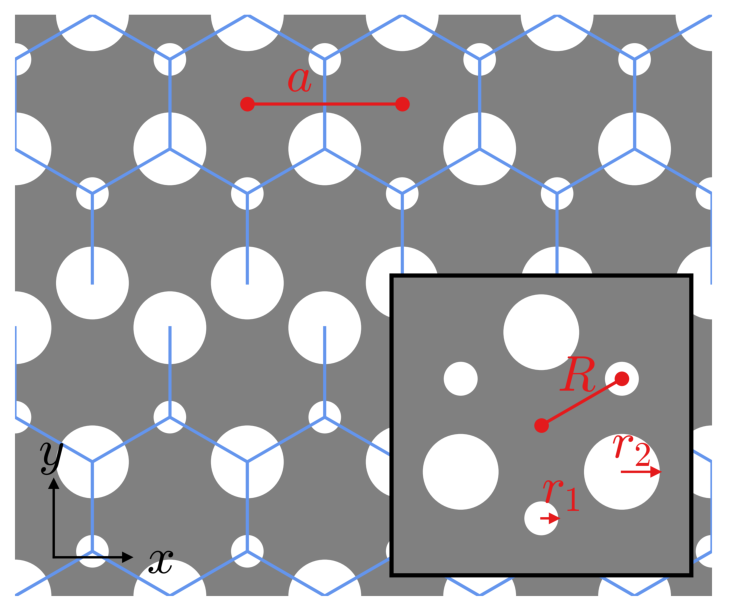}
    \caption{Two schematics of the topological waveguide proposed by He {\em et al.}~\cite{heSilicononinsulatorSlabTopological2019}. The 3D schematic (top) is composed of a Si PCS (blue) on top of a SiO$_2$ substrate (red). The 2D lattice schematic (bottom) has relevant geometric quantities labeled in red; the air holes are shown in white and the direction of propagation is designated as the $x$ direction.}
    \label{fig: Fig 1}
\end{figure}

We start our inverse design optimization from a state-of-the-art topological PCS waveguide design recently introduced by He {\em  et al.}~\cite{heSilicononinsulatorSlabTopological2019}. As shown in Fig. \ref{fig: Fig 1}, this lattice structure is formed from two photonic crystals (PCs), which can be described as a triangular lattice of unit cells containing two air holes, with lattice constant $a$ (see Fig. \ref{fig: Fig 1}), or alternatively as a triangular lattice of 6 hole honeycomb unit cells with lattice constant $\sqrt{3}a$ and cluster radius fixed at $R=a/\sqrt{3}$. In each two-hole unit cell, the two circular air holes have different radii, $r_1$ and $r_2$. Above the interface, the relative positions of the smaller and bigger holes are flipped about the $x$ axis. Both PCs therefore have the same band structure, which contains a photonic bandgap (PBG). We define the PBG as a frequency range below the light line, where there are no PC modes. We use a lattice constant of $a = 453$ nm, $r_1 = 47.7$ nm, $r_2 = 106.5$ nm, and the slab height is $h=4a/7 = 258.9$ nm.
%We define the PBG as a frequency range below the light line, where there are no radiation modes and intrinsic losses. 

\begin{figure}[t]
    \centering
    \includegraphics[width = \linewidth]{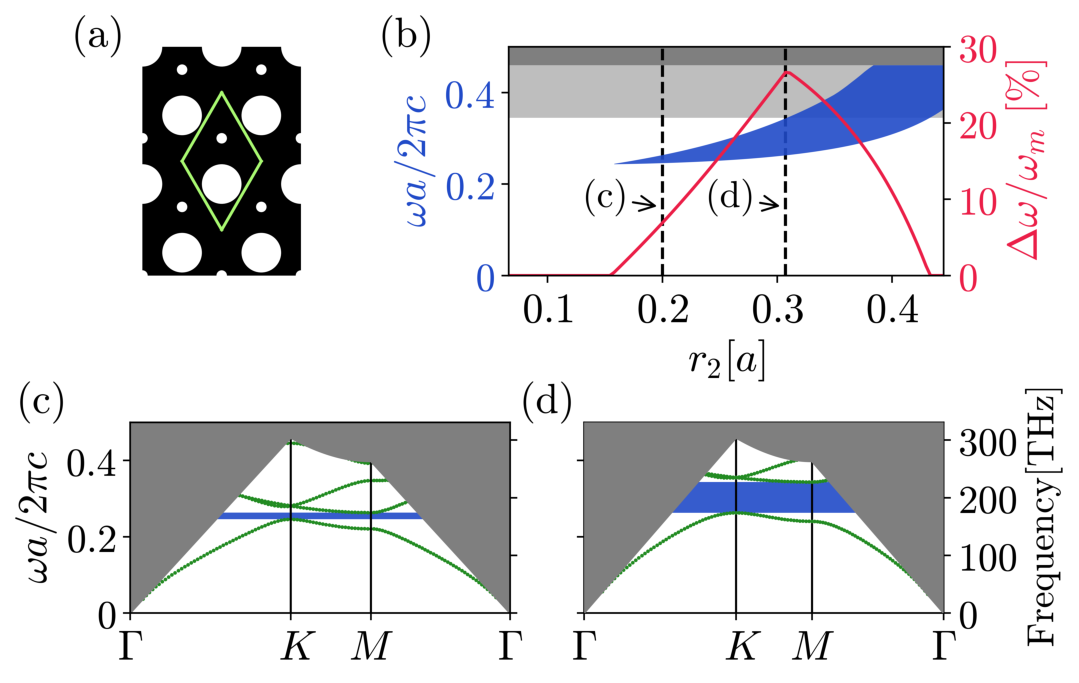}
    \caption{Summary of a methodical search with different values of $r_2$ to increase the band gap, with $r_1 = 30 \, {\rm nm}$, using a slab height $h = 258.9$ nm. (a) Top down view of the PC, modeled with a triangular lattice, the unit cell is outlined in green. (b) The PBG shaded in blue ("Gap map"), sweeping across $r_2$ with $r_1$ held at 30 nm. The gap-midgap ratio is plotted in red, not considering $\omega a /2 \pi c > 0.34$ (shaded in light grey) as part of the PBG. (c) Sample band structure with $r_2$ at the first vertical dashed line in (b). (d) Sample band structure with $r_2$ at the second vertical dashed line in (b).}
    \label{fig: Fig 2}
\end{figure}

%
In Fig.~\ref{fig: Fig 2}, 
we show the PBG properties for various
system paramaters; the PBG is shaded dark blue in panels (b-d).
The interface is constructed by truncating each PC at a value of $y$ at which a row of holes with radius $r_2$ exists and positioning the PCs such that the triangular lattice is continuous across the interface.
To be consistent with He {\em et al.}~\cite{heSilicononinsulatorSlabTopological2019}, the PC and the waveguide are modeled with a silica substrate with $\epsilon = 2.1$, and the slab dielectric constant is $\epsilon = 12.0$.

The objective of the optimization is to increase the bandwidth of a target single guided mode. We define the bandwidth as the frequency range, under the light line, in which no PC bands exist, where the guided mode band exists as a one-to-one transformation of frequency. 
First, the hole sizes in the PC that the waveguide is built from is optimized to increase the size of the PBG, then the interface holes are optimized for the bandwidth. The PC is modeled as a triangular lattice of two-hole cells, which are outlined outlined with a green line in Fig.~\ref{fig: Fig 2}(a). Only the hole radii in each unit cell are allowed to be modified and having assuming a lattice constant, $a$, of 453 nm, are not allowed to fall bellow 30 nm such that the holes can be fabricated with modern fabrication technology \cite{christiansenTopologicalInsulatorsTopology2019}. The holes are also required to maintain 30 nm between their edges, and we keep the slab height at $h = 258.9$ nm. The optimization quickly reduced the smaller hole's radius to the minimum allowed size. At this point, a methodical search across the possible values of $r_2$, with $r_1$ held at 30 nm, is conducted. At each hole size, the band structure and the PBG are calculated. In Fig.~\ref{fig: Fig 2}(b),  the PBG range across values of $r_2$ is shaded in dark blue. In Fig. \ref{fig: Fig 2}(c) and (d), sample band structures are shown for the values of $r_2$ indicated by the vertical dashed black lines in Fig.~\ref{fig: Fig 2}(b). Using the GME technique, the methodical search across 50 values of $r_2$ takes less than 5 minutes to run on a laptop. From the PBG, we calculate the gap-midgap ratio, $\Delta \omega / \omega_m$, which we define as the PBG width divided by its central frequency value. For this calculation, the region $\omega a /2 \pi c > 0.34$ (shaded in light grey in Fig. \ref{fig: Fig 2}(b)) was not considered as part of the PBG because this region is above the light line for the waveguide. A radius of approximately $0.31a$ (140 nm) is chosen for the second PC hole radius.

\begin{figure}[t]
    \centering
    \includegraphics[width=\linewidth]{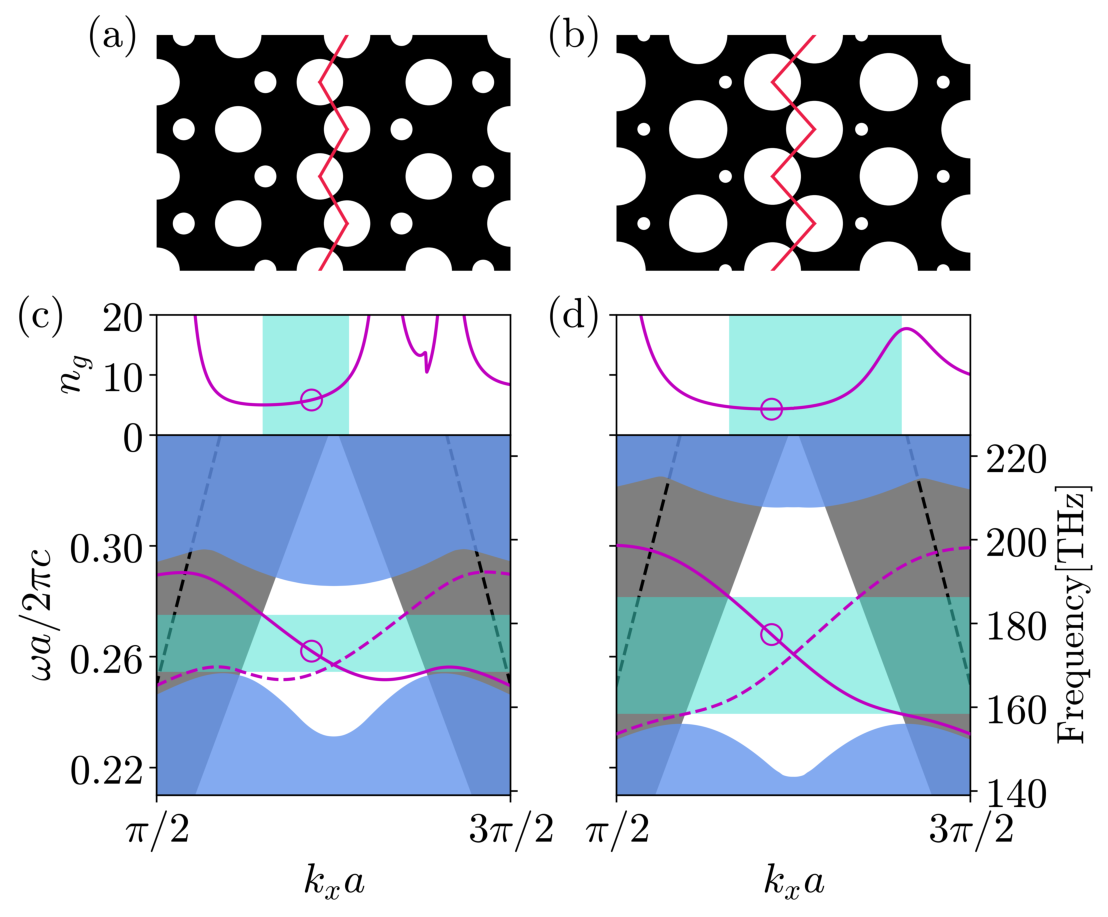}
    \caption{(a) Schematic (top down view) of the initial design with the interface holes indicated by the red line. (b) Schematic of the new design. (c) Group index, $n_g = |c/v_g|$ and band structure for the initial design. The light line for the silica substrate in shaded in grey, the light line for an air bridge is the dashed black line. The PC bands are shaded in light blue, the bandwidth is shaded turquoise, and the two guided bands are plotted in magenta. All bandwidth and $n_g$ calculations are performed for the guided band drawn with a solid line. (d) Group index and band structure for the new design.}
    \label{fig: Fig 3}
\end{figure}

\begin{figure}[t]
    \centering
\includegraphics[width=\linewidth]{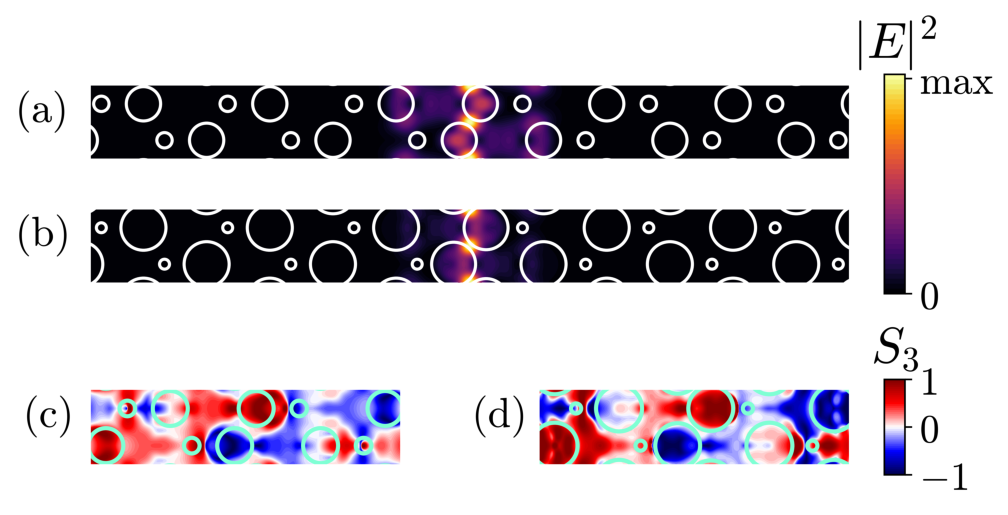}
    \caption{(a) Zoom in of the Bloch mode intensity $|\bm E|^2$ for the initial design; calculations performed for a super cell approximately $34a$ long. Cross sections are taken at the center of the slab height. The field is plotted for the mode and wave vector $\bm k$ indicated by the circular marker in Fig.~\ref{fig: Fig 3}(c), at $k_x a = 15\pi/16$. (b) Bloch mode intensity for the new design. (c) The associated $S_3$ plot for the initial design, further zoomed in to around the waveguide interface (d) Associated $S_3$ plot for the new design.}
    \label{fig: Fig 4}
\end{figure}

This PC is then used to form a waveguide (cf.~\cite{heSilicononinsulatorSlabTopological2019}) as shown in Fig.~\ref{fig: Fig 1}, except the interface hole radii are chosen to be smaller than $r_2$ to increase the initial distance between the edges of the interface holes. For our purposes, using the GME technique, the waveguide is modeled with a super cell $a$ wide in the $x$ direction. As the computation time increases with the size of the super cell, we chose a length of approximately $13a$ in the $y$ direction for the optimization calculations. To obtain more accurate calculations of the relevant PC bands which define the PBG, such as those performed for Figs.~\ref{fig: Fig 3} and \ref{fig: Fig 4}, a super cell approximately $30a$ long in the $y$ direction is used. The super cell with width $a$ contains two holes which form the interface. The two interface holes' radii and position are allowed to be modified as well as the slab height, while enforcing the same geometric constraints as described above for the PC. The FOM used is the bandwidth-mid bandwidth ratio, which we define as the bandwidth divided by its central frequency value. We then used the stochastic optimization algorithm termed ``Adam''~\cite{kingmaAdamMethodStochastic2017} with an objective function returning this FOM. 

Figure \ref{fig: Fig 3} presents schematics of the initial and final designs in (a) and (b) and their respective band structures in (c) and (d), respectively.
The initial design has a bandwidth of 169.4 to 182.7 THz (1.770 to 1.641 $\mu$m) while the new design has a bandwidth of 159.1 to 187.1 THz (1.885 to 1.603 $\mu$m).
Thus, we have improved the operation bandwidth for bound mode
operation by more than 110\%, and the general properties of the mode are also better confined within the PBG.  As shown in Fig.~\ref{fig: Fig 3}, both designs have a similar $n_g$ dispersion curve, though the new design has a region of slightly larger $n_g$ near the bandwidth edge, and far better dispersive properties in momentum space.

Carrying out 10 optimization iterations takes approximately 25 minutes to run on a standard computer workstation. The presented final design is the $347_{th}$ iteration; the optimization took approximately 14 hours to run, which is quite remarkable given the complexity of solving a full 3D complex design problem.
%in nanophotonics.
%
The new design is specified as follows: continuing to use $a = 453$ nm, the optimized slab height is 477 nm; the two interface holes' $x$ coordinate have each been changed by less than 0.1 nm; and the two interface holes' $y$ coordinate have each been modified by $35.4$ nm, such that they are both moved away from the center of the interface. The holes' radii are 136.5 and 136.7 nm.

Finally, we study the spatial details of the Bloch modes.
The polarization of the electric field $\bm E(\bm r)$ can be described by the four Stokes parameters \cite{langStabilityPolarizationSingularities2015}, and we study the normalized third Stokes parameter to quantify the local chiral nature of the modes,
\begin{equation}
    \label{eq: S3}
    S_3(\bm r) = \frac{2 \operatorname{Im}[E_x^*(\bm r) E_y(\bm r)]}{|E_x(\bm r)|^2 + |E_y(\bm r)|^2},
\end{equation}
where points at which $S_3(\bm r)$ is equal to $\pm 1$ are points where the electric field is right- or left-circularly polarized, respectively.

In Fig.~\ref{fig: Fig 4}, cross-sections of the normalized electric field Bloch mode intensity $|\bm E(\bm r)|^2$, as well as the associated $S_3(\bm r)$, are plotted for the initial design in (a) and (c), and the new design in (b) and (d); these plots are all taken at the center of the slab height. The Bloch modes show that the field is well confined to the interface for both designs, and there is reasonable overlap with the chiral points, which can be exploited for unidirectional single photon 
emission~\cite{young_polarization_2015,sollnerDeterministicPhotonEmitter2015, JalaliMehrabad2020}. The Purcell factor from
a quantum emitter at position
${\bf r}_0$, with dipole direction
${\bf n}_d$, 
is proportional to 
%$n_g  ({\bf n}_d^\dagger \cdot {\bf e}_k({\bf r}_0) {\bf e}_k({\bf r}_0) \cdot {\bf n}_d)$,
 $n_g  ({\bf n}_d^\dagger \cdot [\bm E(\bm r_0)  \bm E^*(\bm r_0) +  \bm E^*(\bm r_0)  \bm E(\bm r_0)] \cdot {\bf n}_d)$.
with $\int_{\rm cell} d \bm r \ \epsilon(\bm r) |\bm e_k(\bm r)|^2 =1$.
%which can also be written in terms of 
%an effective mode volume.
Without any additional optimization,
our improved design already yields
more than a factor of 6 improvement 
of the best chiral point Purcell factor, which also accounts for the 
increased maximum $n_g$ below the light line  (from 9.4 to 17.5).

% $n_g / V_{\rm eff}$ \cite{manga_rao_single_2007}, defining the effective mode volume as,
% \begin{equation}
%     V_{\rm eff} = \frac{\int_{\rm cell} d \bm r \ \epsilon(\bm r) |\bm E(\bm r)|^2 }{ \epsilon(\bm r_{\rm antinode}) |\bm E(\bm r_{\rm antinode})|^2 }.    
% \end{equation}
% where the integral is performed over a unit cell of the PCS. The initial design's maximum $n_g$ within the bandwidth is 9.4, for the new design it is 17.5. As an example, $V_{\rm eff}$ for the initial and new designs, for the modes in Fig.~\ref{fig: Fig 4}, is equal to $0.013 \mu {\rm m}^3$ and $0.010 \mu {\rm m}^3$ respectively. If in calculating $V_{\rm eff}$, instead of using $\bm r_{\rm antinode}$, we only consider positions where the electric field is circularly polarized, using $|S_3| > 0.999$ as our threshold, then the minimum $V_{\rm eff}$ is $0.303 \mu {\rm m}^3$ for the initial design and $0.089 \mu {\rm m}^3$ for the new design.}

\section{Conclusions}
Using an efficient computation method to obtain
PCS waveguide band structures (GME) and automatic differentiation, we have
achieved a significant improvement to a state-of-the-art design. The same techniques can be applied to optimize various PCS devices for a variety of different objectives such as quantum dot coupling at a chiral point, achieving extremely slow light, etc. In this work, the chosen device parameterization significantly constrained the optimization to only modify the size of the PC holes and the size and position of holes which define the interface. Increasing the degrees of freedom while still using a shape parameterization by, e.g., modifying the position of the holes in each PCS super cell, allowing the shape of the holes to change, or allowing more holes to be modified while optimizing the interface, should allow even greater improvements to be obtained, while still maintaining the ease of using a shape parameterization.

\section{Funding}
Canadian Foundation for Innovation; 
the Natural Sciences and Engineering Research Council of Canada;
and Queen's University, Canada.

 \section{Acknowledgments}
 We thank Juan Pablo Vasco and Nir Rotenberg for useful discussions.

\bibliography{references, ref2}

\end{document}